\documentclass[%
 reprint,
superscriptaddress,
 amsmath,amssymb,
 aps,
 prl,
]{revtex4-2}

\usepackage{graphicx}
\usepackage{dcolumn}
\usepackage{bm}
\usepackage{xcolor}
\usepackage{hyperref}
\usepackage{tikz}
\usepackage{amsmath}
\usepackage{lipsum} 
\usepackage{graphicx} 
\usepackage{environ} 
\usepackage{braket}

\begin{document}

\preprint{APS/123-QED}

\title{The Bose-Marletto-Vedral proposal in different frames of reference and the quantum nature of gravity
}

\author{Antonia Weber}
\email{antweber@ethz.ch}
\affiliation{Clarendon Laboratory, University of Oxford, Parks Road, Oxford OX1 3PU, United Kingdom}
 \affiliation{Institute for Theoretical Physics, ETH Zurich, 8093 Zurich, Switzerland}
 
\author{Vlatko Vedral}%
 \email{v.vedral@physics.ox.ac.uk}
\affiliation{Clarendon Laboratory, University of Oxford, Parks Road, Oxford OX1 3PU, United Kingdom}

\date{\today}

\begin{abstract}
Observing spatial entanglement in the Bose-Marletto-Vedral (BMV) experiment would demonstrate the existence of non-classical properties of the gravitational field. We show that the special relativistic invariance of the linear regime of general relativity implies that all the components of the gravitational potential must be non-classical. This is simply necessary in order to describe the BMV entanglement consistently across different inertial frames of reference. On the other hand, we show that the entanglement in accelerated frames could differ from that in stationary frames.   
\end{abstract}

\maketitle

\begin{figure}
\begin{tikzpicture}

\draw[fill=white] (1,3) rectangle (5,4);
\draw (1,3) -- (0,3);
\draw (5,4) -- (6,4);

\draw[fill=white] (1,1) rectangle (5,2);
\draw (1,1) -- (0,1);
\draw (5,2) -- (6,2);

\draw[ <->] (2.5,1) -- (2.5,3);
\draw (2.3,2.7) node[below,] { $d_1$};

\draw[ <->] (3.5,1) -- (3.5,4);
\draw (3.3,2.7) node[below,] { $d_2$};

\draw[thick] (0.75,1.25) -- (1.25,0.75);
\draw[thick] (0.75,3.25) -- (1.25,2.75);
\draw[thick] (4.75,2.25) -- (5.25,1.75);
\draw[thick] (4.75,4.25) -- (5.25,3.75);
\draw(0.2,1) circle (2.5pt);
\draw(0.2,3) circle (2.5pt);
\node at (0.2,0.7) {$m_2$};
\node at (0.2,2.7) {$m_1$};

\draw (1.8,1.4) node[below,] { $l_2$};
\draw (1.8,3.4) node[below,] { $l_1$};
\draw (1.8,2.3) node[below,] { $u_2$};
\draw (1.8,4.3) node[below,] { $u_1$};

\end{tikzpicture}
    \caption{The BMV setup. Two masses are put in a spacial superposition of two locations each in two Mach-Zehnder interferometers. They interact only via the gravitational field during an interaction time $\tau$.}
    \label{fig.setup}
\end{figure}
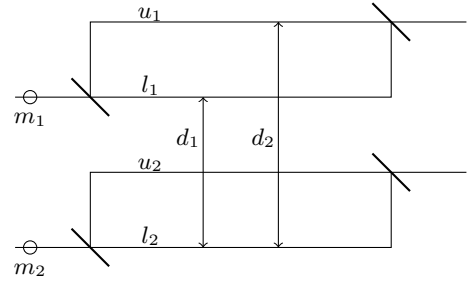

\textit{Introduction}. The problem of quantizing the gravitational field is one of the biggest outstanding questions in physics. A direct detection of the graviton, the hypothetical quantum elementary particle responsible for mediating gravity, would constitute a definitive proof of the need to quantize general relativity. However, current experiments are not yet advanced enough to detect gravitons due to the weakness of gravitational forces \cite{Dyson}. This is in stark contrast with the state of the art in quantum optics where tomography of quantum states of light is routine \cite{Cramer}. The reason, of course, is the weakness of the gravitational force. 
\\
The BMV experiment \cite{MarlettoVedral, Bose} is a recent proposal for a table-top test that would address the question of the quantum nature of gravity, but without the need for energies required to detect gravitons. Instead, two equal masses $m_i$ with $i=1,2$ are each put in a spatial superposition of two locations. This can be done using two Mach-Zehnder interferometers with the lower branch $l_i$ and upper branch the $u_i$ (See Figure \ref{fig.setup}). The distance between both lower branches of the interferometers is $d_1$ while between the lower and the upper branch the distance is $d_2$. Each mass is then held in a superposition of two locations $\frac{1}{\sqrt{2}}(\ket{l}_i+\ket{u}_i)$ during an interaction time $\tau$ and the two masses are assumed to be coupled to each other only via the gravitational field. If gravity in the linear regime behaves like the electromagnetic field, this quantum superposition would acquire a different phase for each possible superposition of paths of the two masses. The resulting simple evolution would then lead to an entangled state, whose amount of entanglement would be a function of the relative phase $\phi$.
\\
Obtaining entanglement in the BMV experiment is thus an indirect witness of the quantum nature of gravity. It implies that the mediating field, which is assumed to couple locally to each mass, must have at least two non-commuting variables, defining it as a quantum system \cite{MarlettoVedral, constructor}. The reason is simply that, since the masses do not couple directly, the only way that they could become entangled is through the gravitational field.

In this work, we show that a simple argument based on special relativity leads us to conclude that the successful BMV experiment implies more than one quantum degree of freedom in the gravitational field. Specifically, we use the fact that the amount of entanglement in different inertial frames in the weak gravitational field must be the same to conclude that not only the scalar mode of gravity has to be quantum but also its vector modes. 

Our derivation follows the analogous results in electromagnetism. The analogy between linearized gravity and electromagnetism is well known (see e.g., \textit{gravitomagnetism} \cite{mashhoon}). We show that in order to obtain a \textit{complete} account of the BMV effect in the linear regime, the resulting entanglement has to be consistently described across different inertial frames. This is only possible if all the 4 components of the gravitational potential, in the Lorenz gauge, are quantum mechanical. 

Finally, we show that, somewhat surprisingly, entanglement is no longer preserved in accelerated frames of reference. In an adapted version of the so-called Bell paradox \cite{Bell, petkov}, we demonstrate that entanglement in the BMV experiment can be lower for a special case of accelerated frames compared to the stationary frames of reference. 
\\

\textit{Background}. In the regime of linearized gravity, the gravitational field is weak such that it can be described by flat Minkowski metric $\eta_{\mu\nu}$ with a small first-order perturbation $|h_{\mu\nu}| \ll 1$: 
\begin{equation}
    g_{\mu\nu}=\eta_{\mu\nu}+h_{\mu\nu} \; .
\end{equation}
In first order perturbation theory, Einstein’s field equation in the Lorenz gauge take the form \cite{carroll}:
\begin{equation}
    \Box\overline{h}_{\mu\nu}= \frac{-16\pi G}{c^4} T_{\mu\nu}\; , 
    \label{eq.einsteq}
\end{equation}
where $\overline{h}_{\mu\nu}=h_{\mu\nu}-\frac{1}{2}\eta
_{\mu\nu}h$ is the trace-reversed metric, $h=\eta_{\mu\nu}h^{\mu\nu}$ and $T_{\mu\nu}$ is the stress-energy tensor.

Equation \ref{eq.einsteq} resembles the well-known relativistic Maxwell equation for the four potential $A^{\mu}$ in the Lorenz gauge \cite{Christodoulou}:
\begin{equation}
    \Box A^{\mu}= \frac{-4\pi k }{ c^2} j^{\mu}\; ,
    \label{eq.relme}
\end{equation}
where $j^{\mu}$ is the four current and $k$ is the Coulomb constant. The solution to equation \ref{eq.relme} is given by the Liénard-Wiechert potential \cite{Jackson}

\begin{equation}
        A^{\mu}= \frac{k}{c^2}\int d^3x' \frac{j^{\mu}(\Vec{x}', t_r)}{|\Vec{x}-\Vec{x}'|}
    \label{eq:lienard}
 \end{equation}
This is a function of the retarded time $t_r=t-\frac{|\vec{x}'-\vec{x}|}{c}$ which simply accounts for the causal propagation of disturbances. 

Following the analogy between linear gravity and electromagnetism, the solution to Einstein’s field equations \ref{eq.einsteq} is given by \cite{carroll}:

\begin{equation}
\overline{h}_{\mu\nu}=\frac{4G}{c^4} \int d^3x' \frac{T_{\mu\nu}(\vec{x}',t_{r})}{|\vec{x}-\vec{x}'|}
\label{eq:soleeq}
\end{equation} 

For slowly moving matter, $|v| \ll c$, the purely spatial components $\overline{h_{ij}}$ can be neglected as they are of order $O({c^{-4}})$ \cite{mashhoon}. In the following analysis, we only consider terms up to this order.

Analogous to electromagnetism, the metric perturbation in linear gravity has an (electric-like) scalar mode $h_{00}$ and (magnetic-like) vector mode $h_{i0}$ for $i=1,2,3$. For more details on gravitomagnetism, see for example \cite{mashhoon}.
\\

This treatment so far is classical. The simplest way to quantize it - and we will see that this is what the BMV experiment analysed in different frames forces us to do - is to upgrade both $|h_{\mu\nu}| \ll 1$ and $T_{\mu\nu}$ into operators \cite{vedral}. The superposed mass in the BMV experiment forces us to regard components of $T_{\mu\nu}$ as operators, while the resulting relativistic invariance of entanglement will imply that the components of $|h_{\mu\nu}| \ll 1$ must also be operators: 

\begin{equation}
\overline{\hat{h}}_{\mu\nu}=\frac{4G}{c^4} \int d^3x' \frac{\hat{T}_{\mu\nu}(\vec{x}',t_{r})}{|\vec{x}-\vec{x}'|}\; .
\label{eq:soleeqq}
\end{equation}

The metric perturbation written in terms of the graviton creation $a^\dagger$ and annihilation $a$ operators in the Fourier space is \cite{when, Boughn}
 \begin{equation}
 \begin{aligned}
 \hat{h}_{\mu\nu} \propto & \int d^3 k \frac{1}{\sqrt{\omega_k}} \left[ a(k) e^{ik_\nu x^\nu}
+a(k)^\dagger e^{-ik_\nu x^\nu} \right]
\label{eq.quantization}
      \end{aligned}
 \end{equation}
where $k$ is the wave number, $\omega$ is the frequency of the mode. We assume a single polarization for simplicity and without any loss of generality.
\\

\textit{The BMV proposal in a stationary frame of reference.}
In a stationary frame of reference $K$ the energy momentum tensor for two masses $m:=m_1=m_2$ described as point particles at locations $\vec x_1$ and $\vec x_2$ is
\begin{equation}
    T^{00}(\vec x, t)=mc^2(\delta(\vec{x}-\vec{x}_1(t))+\delta(\vec{x}-\vec{x}_2(t)))
    \label{eq:energymomentum}
\end{equation} and $T^{\mu\nu}(\vec x, t)=0$ otherwise.
We quantize this by turning the stress-energy tensor in an operator: $T_{00} \rightarrow \hat{T}_{00} \propto b^\dagger b + c^\dagger c $. Here $b^\dagger, c^\dagger, b, c$ are the respective creation and annihilation operators of the two masses. This is necessary due to the superposition of locations of the masses in the interferometers. 
Each mass then couples to the respective other mass through the interaction Hamiltonian density \cite{Boughn} $\mathcal{H}_{int}=-\frac{1}{2}h_{\mu\nu}T^{\mu\nu}$. Hence, it is also necessary to quantize $h_{\mu\nu}$ \ref{eq:soleeqq}.

For stationary observers, only the $00$-component of the stress-energy tensor \ref{eq:energymomentum} is non-vanishing. Hence, we have to upgrade the scalar mode of gravity to a quantum degree of freedom i.e., $h_{00} \rightarrow \hat{h}_{00}$.

Let us now examine more closely the non-commuting variables that lead to the quantum nature of the gravitational field. 
This is evident by considering the total Hamiltonian, which consists of the free Hamiltonian of the field, the free Hamiltonian of the masses and the interaction Hamiltonian \cite{when, Boughn}:
\begin{equation}
\begin{aligned}
        \hat{H}&= \sum_k  \hbar \omega_k a_k^\dagger a_k + mc^2 (b^\dagger b+ c^\dagger c) \\
        &\quad + \sum_k \frac{\hbar g_k}{\sqrt{\omega_k}} b^\dagger b  \left[ a_k e^{ikx_1}+a_k^\dagger e^{-ikx_1}\right] \\
        &\quad + \sum_k \frac{\hbar g_k}{\sqrt{\omega_k}} c^\dagger c \left[ a_k e^{ikx_2}+a_k^\dagger e^{-ikx_2}\right]
\end{aligned}
\end{equation}
where $g_k=mc \sqrt{\frac{2\pi G}{\hbar \omega_k V}}$ is the gravity-matter coupling constant. We here only consider the relevant modes within the quantization volume $V$\cite{when}. 
It is precisely this non-commutativity of entities - each an infinite sum of modes, with each mode having at least two non-commuting operators of the form $a^\dagger a $ and $a + a^\dagger$- that demonstrates the quantum nature of the mediating gravitational field.
\\
Keeping this in mind, we proceed to calculate the phase $\phi$ for different inertial frames using the path integral formalism.  As shown in \cite{Christodoulou}, the phase $\phi$ is proportional to the on-shell action
\begin{equation}
    \phi=\frac{S_{int}}{\hbar}= \frac{1}{4\hbar} \int d^4x h_{\mu\nu} T^{\mu\nu}  
\label{eq.action}
\end{equation}
Hence, for stationary observers this reduces to
\begin{equation}
\begin{aligned}
      \phi= \frac{1}{4\hbar} \int d^4x h_{00}T^{00}
        \label{eq.actionstationary}
\end{aligned}
\end{equation}

\textit{Moving frames of reference and the quantum nature of the gravitational vector potential}. In the following, we describe the entanglement from a moving frame of reference $K'$, where an observer moves with a constant velocity $\vec{v}=v_x$ along the x-direction perpendicular to the path of the masses.

A na\"{i}ve consideration of solely the phase acquired by a Newtonian interaction (lowest order perturbation theory),
\begin{equation}
    \phi=\frac{Gm_1m_2\tau}{\hbar d}
    \label{eq:Newtonian}
\end{equation}
might suggest that the observer in $K'$ measures higher entanglement due to time dilatation and length contraction. However, this cannot be true in a consistent theory, as all outcomes of physical measurements (``clicks"), including entanglement (just a combination of correlations between the clicks), must remain invariant across inertial frames.
\\
It is therefore clear that we cannot just use the scalar gravity to account for the BMV experiment. By applying a similar argument within the framework of linearized quantum gravity, we can gain insights into the quantum nature of gravitational modes and resolve the question above about the consistency of the theory across inertial frames of reference.
\\
In the moving frame $K'$, the stress-energy tensor gains non-vanishing spatial quantum degrees of freedom $T_{11}$ and mixed quantum components $T_{01}=T_{10}$ as a result of the Lorentz boost:
\begin{align} T^{0'0'}&=\gamma^2T^{00}  & T^{0'1'}&=-\beta\gamma^2T^{00} \\
T^{1'0'}& =-\beta\gamma^2T^{00} &
T^{1'1'}&=\gamma^2\beta^2T^{00}
\end{align}
        
 $\beta=\frac{v_x}{c}$ and $\gamma=\frac{1}{\sqrt{1-\beta^2}}$ is the Lorentz factor. It follows that in $K’$, $ \{ \hat{T}^{00}, \hat{T}^{01}, \hat{T}^{11} \}$ are all operators.

Assume that only the scalar mode of gravity has a quantum degree of freedom $\hat{h}_{00}$ as above. We calculate the entanglement, proportional to the acquired phase $\phi$, from the moving frame of reference $K'$.  
The scalar part of the metric perturbation transforms under a Lorentz boost as:
\begin{equation}
        h_{0'0'}=\gamma^2h_{00}
\end{equation}

Hence, the phase in $K'$ is proportional to
\begin{equation}
\begin{aligned}
       \phi'&= \frac{1}{4\hbar} \int d^4x' h_{0'0'}T^{0'0'} \\
        &= \frac{1}{4\hbar}\int d^4x 
        \gamma^4h_{00}T^{00} \neq \phi
        \end{aligned}
        \label{eq.action_sc}
\end{equation}
This would imply that two observers in different inertial
frames $K$ and $K$’ measure a different amount of entanglement. 
 As in the case of the Newtonian phase above in equation \ref{eq:Newtonian}, the entanglement between the two masses would be higher in the case of a moving observer.
Again, this cannot be a complete theory as Lorentz invariance is violated for different inertial frames.
\\
Instead, assume that not only the scalar components but also the gravitational vector potential has quantum degrees of freedom i.e., $h_{0i}\rightarrow \hat{h}_{0i}$ for $i=0,1,2,3$. In the moving frame $K'$, they can be expressed in terms of the modes in the stationary frame $K$ as
\begin{align}
    h_{0'0'}&=\gamma^2h_{00} &
    h_{0'1'}&=\beta\gamma^2h_{00} &
    h_{1'0'}&=\beta\gamma^2h_{00}
\end{align}

Hence, the entanglement in $K'$ is proportional to the phase
\begin{equation}
    \begin{aligned}
    \phi' &= \frac{1}{4\hbar} \int d^4x' h_{0'0'}T^{0'0'}+2h_{0'1'}T^{0'1'} \\
    &= \frac{1}{4\hbar} \int d^4x 
        h_{00}T^{00}\gamma^4(1-2\beta^2)
    \end{aligned}
\end{equation}

Keeping terms up to $O(c^{-4})$, we expand $\gamma^4\thickapprox 1+2\beta^2+O(c^{-4})$. 

\begin{equation}
\begin{aligned}
        \phi'&\approx \frac{1}{4\hbar} \int d^4x 
        h_{00}T^{00}(1+2\beta^2)(1-2\beta^2)
    \\ &\approx \frac{1}{4\hbar} \int d^4x 
        h_{00}T^{00} =\phi
        \end{aligned}
\end{equation}

This leads to the same account of entanglement as in the stationary case.
Entanglement can only be consistently described across inertial frames if 
not only the scalar part of the gravitational field $h_{00}$ has a quantum degree of freedom but also the graviomagnetic vector potential $h_{0i}$. This follows from a generalisation of the above argument for all directions of space. All 4-degrees of freedom of the gravitational potential have to be quantum as a necessity to have a complete description within the confines of linearized quantum gravity. We refer the reader to similar arguments for the existence of the gravitational vector potential in the (classical) theory of general relativity \cite{Nordtvedt}.

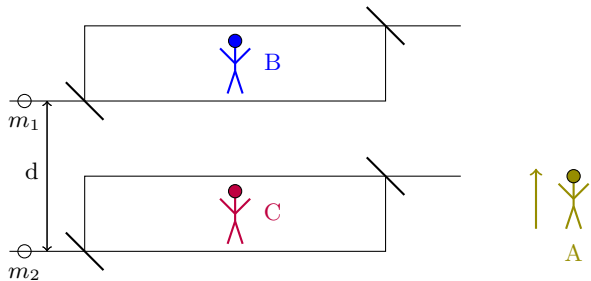
\begin{figure}
\begin{tikzpicture}

\draw[thick, olive] (7.4,1.3) -- (7.5,1.6);
\draw[thick, olive] (7.5,1.6) -- (7.6,1.3);
\draw[thick, olive] (7.5,1.6) -- (7.5,2.0);
\draw[fill=olive] (7.5,2.0) circle (2.5pt);
\draw[thick, olive] (7.5,1.7) -- (7.7,1.9);
\draw[thick, olive] (7.5,1.7) -- (7.3,1.9);
\draw (7.5,1.2) node[below, olive] {A};

\draw[fill=white] (1,3) rectangle (5,4);
\draw (1,3) -- (0,3);
\draw (5,4) -- (6,4);
\draw[thick, blue] (2.9,3.1) -- (3,3.4);
\draw[thick, blue] (3,3.4) -- (3.1,3.1);
\draw[thick, blue] (3,3.4) -- (3,3.8);
\draw[fill=blue] (3,3.8) circle (2.5pt);
\draw[thick, blue] (3,3.5) -- (3.2,3.7);
\draw[thick, blue] (3,3.5) -- (2.8,3.7);
\draw (3.5,3.75) node[below, blue] {B};

\draw[fill=white] (1,1) rectangle (5,2);
\draw (1,1) -- (0,1);
\draw (5,2) -- (6,2);
\draw[thick, purple] (2.9,1.1) -- (3,1.4);
\draw[thick, purple] (3,1.4) -- (3.1,1.1);
\draw[thick, purple] (3,1.4) -- (3,1.8);
\draw[fill=purple] (3,1.8) circle (2.5pt);
\draw[thick, purple] (3,1.5) -- (3.2,1.7);
\draw[thick, purple] (3,1.5) -- (2.8,1.7);
\draw (3.5,1.75) node[below, purple] { C};

\draw[thick, ->, olive] (7,1.3) -- (7,2.1);

\draw[ <->] (0.5,1) -- (0.5,3);

\draw[ <->] (0.5,1) -- (0.5,3);
\draw (0.3,2.3) node[below,] { d};

\draw[thick] (0.75,1.25) -- (1.25,0.75);
\draw[thick] (0.75,3.25) -- (1.25,2.75);
\draw[thick] (4.75,2.25) -- (5.25,1.75);
\draw[thick] (4.75,4.25) -- (5.25,3.75);
\draw(0.2,1) circle (2.5pt);
\draw(0.2,3) circle (2.5pt);
\node at (0.2,0.7) {$m_2$};
\node at (0.2,2.7) {$m_1$};
\end{tikzpicture}
    \caption{An observer $A$ is accelerated with respect to the two interferometers of the BMV experiment. The observers $B$ and $C$ in the interferometers are at rest.}
    \label{fig:acc_obs}
\end{figure}

\begin{figure}
\begin{tikzpicture}

\draw[thick, olive] (7.4,1.3) -- (7.5,1.6);
\draw[thick, olive] (7.5,1.6) -- (7.6,1.3);
\draw[thick, olive] (7.5,1.6) -- (7.5,2.0);
\draw[fill=olive] (7.5,2.0) circle (2.5pt);
\draw[thick, olive] (7.5,1.7) -- (7.7,1.9);
\draw[thick, olive] (7.5,1.7) -- (7.3,1.9);
\draw (7.5,1.2) node[below, olive] {A};

\draw[fill=white] (1,3) rectangle (5,4);
\draw (1,3) -- (0,3);
\draw (5,4) -- (6,4);
\draw[thick, blue] (2.9,3.1) -- (3,3.4);
\draw[thick, blue] (3,3.4) -- (3.1,3.1);
\draw[thick, blue] (3,3.4) -- (3,3.8);
\draw[fill=blue] (3,3.8) circle (2.5pt);
\draw[thick, blue] (3,3.5) -- (3.2,3.7);
\draw[thick, blue] (3,3.5) -- (2.8,3.7);
\draw (3.5,3.75) node[below, blue] {B};

\draw[fill=white] (1,1) rectangle (5,2);
\draw (1,1) -- (0,1);
\draw (5,2) -- (6,2);
\draw[thick, purple] (2.9,1.1) -- (3,1.4);
\draw[thick, purple] (3,1.4) -- (3.1,1.1);
\draw[thick, purple] (3,1.4) -- (3,1.8);
\draw[fill=purple] (3,1.8) circle (2.5pt);
\draw[thick, purple] (3,1.5) -- (3.2,1.7);
\draw[thick, purple] (3,1.5) -- (2.8,1.7);
\draw (3.5,1.75) node[below, purple] { C};

\draw[thick, ->, blue] (2.5,3.1) -- (2.5,3.9);
\draw[thick, ->, purple] (2.5,1.1) -- (2.5,1.9);

\draw[ <->] (0.5,1) -- (0.5,3);
\draw (0.3,2.3) node[below,] { d};

\draw[thick] (0.75,1.25) -- (1.25,0.75);
\draw[thick] (0.75,3.25) -- (1.25,2.75);
\draw[thick] (4.75,2.25) -- (5.25,1.75);
\draw[thick] (4.75,4.25) -- (5.25,3.75);
\draw(0.2,1) circle (2.5pt);
\draw(0.2,3) circle (2.5pt);
\node at (0.2,0.7) {$m_2$};
\node at (0.2,2.7) {$m_1$};


\end{tikzpicture}
    \caption{The two interferometers of the BMV experiment are accelerated at the same time to the same degree, the observer $A$ in the lab frame is at rest.   }
    \label{fig:accel_interf}
\end{figure}
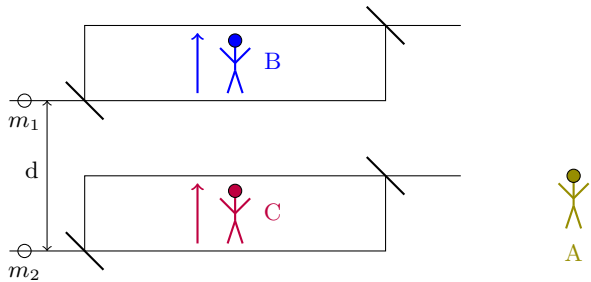

\textit{Accelerating frames of reference and adapted Bell’s paradox}. Finally we comment on the fact that observers can measure a different amount of entanglement when it comes to accelerated motion. In the following, we discuss the difference between an account of entanglement in the BMV experiment when \textit{(1)} the observer $A$ is accelerated (see Figure \ref{fig:acc_obs}), and \textit{(2)} the two interferometers are simultaneously equally accelerated, while leaving the lab frame at rest as an adaptation to \textit{Bell’s paradox} \cite{Bell, petkov},   (see Figure \ref{fig:accel_interf}).
\\
Consider observer $A$ who is in accelerated motion with respect to the two interferometers of a BMV experiment as in Figure \ref{fig:acc_obs}. The more the speed of $A$ increases, the higher the contraction of the distance between the two interferometers. Hence, the contribution of the gravitomagnetic vector potential becomes increasingly significant. The entanglement described by observer $A$ does not increase compared to when $A$ is not accelerating. This follows from requiring consistency among observers: the two observers $B$ or $C$ in the interferometers are stationary and do not notice any change in entanglement compared to the situation where $A$ is not accelerating. Hence, in this case of acceleration, the entanglement is the same as for a stationary BMV experiment.
\\
This is different if the two interferometers of the BMV experiment are accelerating at the same time to the same degree and observer $A$ is in a stationary frame (See Figure \ref{fig:accel_interf}). 
Observer $A$ measures a ever greater length contraction of the distance $d$ between the interferometers the more the velocity of the two interferometers increases. This seems contradictory, as due to the same acceleration, the two interferometers have exactly the same velocity at every moment in time. Thus, for observer $A$, the two interferometers should maintain a constant separation $d$ from each other. 

Analogous to Bell’s analysis \cite{Bell, petkov}, the solution to this apparent paradox is that the distance between the two interferometers must increase for $B$ and $C$ exactly as much as it decreases for A due to length contraction. This increase in distance between the interferometers precisely compensates for the decrease due to length contraction, ensuring that there is no paradox. For observer $A$, the distance between the interferometers remains constant at the end, whereas for observers $B$ or $C$, the distance between the interferometers $d$ must increase by a factor of $\gamma$: $d'=d\gamma$.
\\
Hence, as observers $B$ and $C$ are in a stationary frame in which only the scalar part of gravity is relevant, they measure a lower entanglement. This due to the increased distance between the interferometers
   $ \phi'=\frac{Gm_1m_2\tau}{d'}= \frac{1}{\gamma}\phi $
compared to the case where they are not accelerated.  This phenomenon has no classical analogue.
\\
A more detailed discussion of accelerated observers would require the framework of quantum field theory in curved spacetime \cite{Wald}. However, it is worth bearing in mind that quantum field theory in spacetime is a semi-classical theory which would be ruled out if entanglement was measured in the BMV experiment \cite{when}. Nevertheless, it would be interesting to analyze the Bell scenario within this kind of semi-classical perturbative model. 
\\

\textit{Conclusion}. We conclude that in the complete account of linearized quantum gravity the scalar and vector components of the metric perturbation must have quantum degrees of freedom in the sense that they must have at least two sets of non-commuting variables each. Our arguments rested on the fact that different inertial observers have to measure the same amount of entanglement in the BMV experiments. We then show that accelerating the two interferometers to the same degree at the same time results in observers measuring different amounts of entanglement compared to the stationary case. There is of course no contradiction here since this case lies outside of special relativity and is consistent with the account given by general relativity. 

Given that Einstein’s equations are non-linear, however, it would be interesting to analyse the effects of non-linearities on the entanglement in the BMV experiment. These would be higher order terms compared to the ones investigated here and would therefore be beyond the current experimental reach. Nevertheless, it would be important to study their impact on the gravitational spatial superpositions. 

Furthermore, we believe that an analysis of a potential link between the phenomenon of entanglement invariance across different reference frames and the timeless framework of Page-Wootters \cite{pagewootters} could shed more light on the role of time in the BMV proposal. Above all of course we hope for rapid progress of experimental realisations of the BMV and related proposals. It is precisely the lack of any experimental input that has impacted the field of quantum gravity and is responsible for the lack of breakthroughs. 
\newpage
{\bf Acknowledgements}. V.V. thanks the Gordon and Betty Moore Foundation and the Templeton Foundation for supporting his research. We thank Giuseppe Di Pietra for helpful comments.

\bibliography{ref}

\end{document}